\documentclass[journal]{IEEEtran}
\IEEEoverridecommandlockouts
\usepackage{cite}
\usepackage{amsmath,amssymb,amsfonts}
\usepackage{algorithmic}
\usepackage{graphicx}
\usepackage{textcomp}
\usepackage{xcolor}
\usepackage{color}
\usepackage{setspace}

\def\BibTeX{{\rm B\kern-.05em{\sc i\kern-.025em b}\kern-.08em
    T\kern-.1667em\lower.7ex\hbox{E}\kern-.125emX}}
\begin{document}
 
\title{Power Control for ISI Mitigation in Mobile Molecular Communication\\
\thanks{The work was supported by the Natural Science Foundation of China (61271276), and PhD student short-term study abroad project of Xidian University.}}

\author{
\IEEEauthorblockN{Dongliang Jing$^{1,2}$, Yongzhao Li$^1$$^*$ \IEEEmembership{Senior Member,~IEEE}, Andrew W. Eckford$^2$ \IEEEmembership{Senior Member,~IEEE}}

\IEEEauthorblockA{$^1$State Key Laboratory of Integrated Services Networks,
Xidian University, China, Xi'an, 710071\\
$^2$ Department of Electrical Engineering and Computer Science, York University, Toronto, Ontario, Canada}
}
\vspace{-0.5cm}
\maketitle

\begin{abstract}
In mobile molecular communication (MC), inter-symbol interference (ISI) 
can be mitigated by power control, requiring accurate estimates of the distance from transmitter to receiver. We present two power control strategies based on binary concentration shift keying (BCSK), namely BCSK with power control based on distance (BCSK-d), and BCSK with power control jointly considering distance and residual molecules in the channel (BCSK-d-RM). Performance of BCSK-d and BCSK-d-RM are analyzed in terms of the bit error rate (BER), the average energy consumption per bit, and the optimal detection threshold. Simulation results show that BCSK-d-RM and BCSK-d outperform BCSK in BER with the varying distance and the number of transmitted molecules. As BCSK-d-RM has higher performance, while BCSK-d has lower complexity, these schemes present a useful design tradeoff for mobile MC.
\end{abstract}


\section{Introduction}
 \IEEEPARstart{M}{olecular} communication (MC) has emerged as a promising communication approach for nanonetworks. Operated by exchanging chemical signals, it has attracted significant attention due to its potential applications in nano-medicine and nano-sensing\cite{farsad2016comprehensive,akyildiz2019moving}. In MC, the information can be encoded in the number of molecules, the release time of molecules and the type of molecules \cite{haselmayr2017transposition}. 

In diffusion-based molecular communication (DBMC), the released molecules migrate from a higher concentration region to a lower concentration region following Fick's law. During propagation, as distance increases, the proportion of molecules absorbed by the receiver decreases in a bit interval, leading to a larger number of molecules whose arrival at the receiver is delayed; this results in inter-symbol interference (ISI). Therefore, it is challenging to ensure reliability for MC when the distance between transmitter and receiver is variable.

Mobile MC has been studied in recent work  \cite{jamali2019channel,iwasaki2017graph,huang2018mean,varshney2018diffusive}. In \cite{jamali2019channel}, the authors provides the channel models for time-varying MC systems with moving transmitters and receivers for the applications such as smart drug delivery. 
A graph-based model of mobile molecular communication systems to study the spatial-temporal dynamics of bio-nanomachine concentration in a complex environment was developed in \cite{iwasaki2017graph}. Closed-form expressions for the mean and the variance of the received signal by considering the randomness of both the mobility of nanomachines were considered in \cite{huang2018mean}. With a fixed molecule budget at the transmitter
nanomachine, it was shown in \cite{varshney2018diffusive} that the performance of mobile diffusive molecular communication can be significantly enhanced by allocating a larger fraction of the molecules for transmission in later time slots.

In mobile MC, ISI is notably affected by the variation of distance.
In \cite{sun2016adaptive}, the authors proposed an adaptive code width (ACW) protocol to mitigate the ISI in mobile MC. In the ACW protocol, the transmitter adapts the code width based on the measured distance.  In \cite{chang2017adaptive}, an adaptive ISI mitigation method and two adaptive detection schemes are proposed for this mobile MC. 
In the proposed scheme, adaptive ISI mitigation, estimation of dynamic distance, and the corresponding impulse response reconstruction are performed in each symbol interval. However, when the distance is long, there is a large number of molecules remaining in the channel beyond the current bit interval and reduces the data rate.

In this letter, the power control strategies for DBMC are presented to improve communication reliability in the distance-varying scenario.  Compared to \cite{tepekule2015isi} which employs a power adjustment strategy by utilizing the residual molecules from the previous symbols the fixed MC, in this paper, the proposed power control schemes control the number of transmitted molecules based on the distance or jointly considering the distance and the residual molecules from the previous symbols in mobile MC. Compared with the detection methods at the receiver, the power control methods at the transmitter save the transmitted molecules. Analytical expressions are derived to characterize the signal power. The main contributions of this letter include:
\begin{itemize} 

\item A lightweight power control strategy (BCSK-d)  is proposed for mobile MC based on the distance between the transmitter and the receiver; and

\item A computationally complex power control strategy (BCSK-d-RM) is proposed for mobile MC jointly considering the distance between the transmitter and the receiver and the residual molecules from former interval.

\end{itemize}
The BSCK-d-RM has advantages over the BSCK-d in terms of the BER and the average energy consumption. Thus, the two algorithms give a system designer a tradeoff between performance and computational complexity.

\section{System Model}
A 1-dimensional mobile MC channel is considered, in which flow-induced diffusion occurs with flow velocity $v$. In terms of mobility, the key modelling assumptions are: 
\begin{itemize}
    \item Similar to \cite{varshney2018flow}, in the considered mobile MC system, the transmitter is fixed, while the receiver is mobile;
    \item The receiver diffuses along with the information molecules, through a medium with flow velocity $v$; and
    \item The transmitter is assumed to be capable of controlling signal power based on the estimated distance. 
\end{itemize}

A pulse of molecules 
is released into the diffusion channel and propagates in the aqueous medium via diffusion with flow, then some of the molecules are observed by the receiver.
We focus on a mobile MC system with binary CSK modulation, i.e., on-off keying (OOK). Let $\{b_k\} = \{b_0, b_1, \ldots, b_k, \ldots\}$ be the binary information sequence, where $b_k\in\{0, 1\}$. 
The transmitter emits an impulse with signal power $N_{tx}$ molecules to represent $b_k=1$, and emits nothing to represent $b_k=0$. Thus, the transmitted molecular signal pulse $s(t)$ can be expressed as
\begin{equation}
\small
s\left( t \right) = \sum\limits_{k = 0}^\infty  {{b_k}{N_{tx}}\delta \left( {t - k{T}} \right)} ,
\end{equation}
where $\delta(t)$ is the Dirac delta function, and
where $T$ represents the symbol interval, which is also the bit interval when the binary OOK modulation scheme is employed.

In our mobile system, the distance $d_k$ between the transmitter and the receiver can be expressed as ${d_k} = {d_0} + {d_B} + {d_v}$, where $d_0$ is the initial distance, $d_B$ is the displacement due to the Brownian motion, and $d_v$ is the displacement due to the flow. The Brownian motion $d_B$ of the receiver during the $k$th bit interval can be modeled by Wiener process, and follows a Gaussian distribution with zero mean and variance ${\sigma_{rx} ^2} = 2{D_{rx}}T$, where $T$ is the bit interval and $D_{rx}$ is the diffusion coefficient of the receiver, while the flow displacement is given by $d_v = kTv$. Therefore, the distance $d_k$ between the transmitter and the receiver follows a Gaussian distribution
${d_k}\sim \mathcal{N}\left( {\overline d_k,\sigma _{k}^2} \right)$, where $\overline d_k = {d_0} + kTv$ and $\sigma _{k}^2 = 2kTD_{rx}$. The PDF of the Euclidean distance ${\tilde d}_k$ is shown in (Eq. 8, \cite{varshney2018flow}).

 \begin{figure}[t!]
  \centering
  \includegraphics[width=0.3\textwidth]{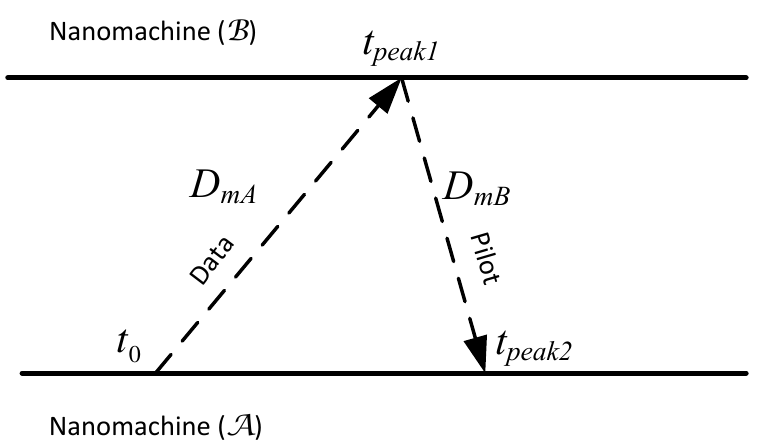}\\
  \caption{The distance estimation process.}
\end{figure}





In MC, distance estimation is usually based on the signal attenuation at the receiver. In this letter, similar to \cite{moore2012measuring}, a nanomachine $(\mathcal{A})$ estimates the distance to another nanomachine $(\mathcal{B})$ by requesting a pilot signal from the nanomachine $(\mathcal{B})$. The nanomachine $(\mathcal{A})$ controls its transmit power based on the estimated distance. This process is shown in Fig. 1. In the first phase of the process, the data is transmitted when the nanomachine $(\mathcal{A})$ emits an impulse with signal power $N_{tx}$ molecules at time $t_0$, then the molecules propagate through the medium. At nanomachine $(\mathcal{B})$, the concentration increases to a peak at time $t_{peak1}$ before decreasing. In the second phase, at the time of $t_{peak1}$, the nanomachine $(\mathcal{B})$ emits a pilot signal of molecules, which encodes $t_{peak1}$. The nanomachine $(\mathcal{A})$ measures the $t_{peak2}$ at the peak of the concentration, and is aware of $t_{peak1}$ which is encoded in the pilot signal. (We do not consider the method of encoding, but we assume that it is error-free.) 

Since the transmitter is stationary while the receiver moves with the flow, the first hitting time distribution is {\em different} for the forward and backward links. The first hitting time for the forward link (transmitter to receiver) follows a L{\'e}vy distribution \cite{varshney2018flow}, since both signal and receiver experience the same drift:
\begin{equation}
\small
f\left( t \right) = \frac{{{{\tilde d}_k}}}{{\sqrt {4\pi {D_\mathrm{eff}}{t^3}} }}\exp \left( { - \frac{{{{\tilde d}_k}^2}}{{4D_\mathrm{eff}t}}} \right),
\end{equation}
where $D_{\mathrm{eff}}=D_m+D_{rx}$ \cite{kim1999exact}, $D_m$ is the diffusion coefficient of the information molecules and $t$  means the relative time of observation of the signaling molecules after the molecules released.

The probability of hitting an absorbing receiver prior to time $t$ can be expressed as
\begin{equation}
\small
P\left( t \right) = \int_0^t {f\left( {t'} \right)} dt' = \mathrm{erfc}\left( {\frac{{{{\tilde d}_k}}}{{\sqrt {4D_{\mathrm{eff}}t} }}} \right),
\end{equation}
where $\mathrm{erfc}\left( x \right) = \frac{2}{{\sqrt \pi  }}\int_x^\infty  {{e^{ - {t^2}}}dt}$ is the complementary error function. The hitting probability during the $k$th bit interval can be simply written as $P_k$.

In the reverse link (receiver to transmitter), the diffusion coefficient of the molecules transmitted from nanomachine $(\mathcal{B})$ is $D_{mB}$.
The solution to the diffusion equation for $N$ molecules released from nanomachine $(\mathcal{B})$ to nanomachine $(\mathcal{A})$ can be written as \cite{farsad2016comprehensive}
\begin{align}
c\left[ {\tilde d,t} \right] = \frac{N}{{\sqrt {4\pi {D_{mB}}t} }}\exp \left( { - \frac{{{{\left( {\tilde d + vt} \right)}^2}}}{{4{D_{mB}}t}}} \right)  
\end{align}

After the emission process, the concentration will peak at a certain time before decreasing. Therefore, when $t = t_{peak2}$,
then $\frac{{\partial c\left[ {{\tilde d}_k,t} \right]}}{{\partial t}} = 0$. It can be shown that, at $t_{peak2}$, the estimate of the distance is given by
\begin{equation}
\small
{\hat d_k} = \sqrt {{v^2}{{\left( {{t_{peak2}} - {t_{peak1}}} \right)}^2} + 2{D_{mB}}\left( {{t_{peak2}} - {t_{peak1}}} \right)}.  
\end{equation}%

%
Due to the flow and Brownian motion, the receiver is still in motion during the $t_{peak2}-t_{peak1}$ time interval. To account for this, we add a distance compensation term, adding the distance during the $t_{peak2}-t_{peak1}$ time interval based on the flow velocity. Therefore, the estimated distance with distance compensation can be expressed as 
\begin{align}
\small
\nonumber
{\hat d_k} &= \sqrt {{v^2}{{\left( {{t_{peak2}} - {t_{peak1}}} \right)}^2} + 2{D_{mB}}\left( {{t_{peak2}} - {t_{peak1}}} \right)}\\
&\:\:\:\:+ v\left( {{t_{peak2}} - {t_{peak1}}} \right).  
\end{align}

Since our main interest is in power control schemes based on distance, we perform distance estimation in every bit interval. As the pilot symbols are in the reverse link, and the data transmission is in the forward link, we do not take these pilot symbols into account in the analysis, instead leaving an analysis of the cost of pilot symbols for future work.


\section{Power control in Mobile Molecular Communication}

For mobile MC, the number of molecules measured at the receiver varies according to the distance, which makes the correct signal judgement more challenging when employing a preset threshold $N_{th}$. Especially, with the increase of distance $d$, the number of received molecules decreases sharply during a bit interval.

In this letter, we propose the power control strategies for the mobile MC. The proposed strategies are to overcome the problem that the number of absorbed molecules changing with the mobility of nanomachines. The power control strategies adjust the transmitted molecules based on the estimated distance to ensure that the number of molecules absorbed by the receiver in a bit interval remains stable for bit 1. Assuming the absorbed molecules in a bit interval remains stable at $N_{rs}$ and if $N_{rs}\geq N_{th}$, a symbol $1$ is decoded, else a symbol $0$ is decoded. Then two power control strategies based on BCSK, namely, BCSK-d and BCSK-d-RM are proposed. Finally, the computational complexity of the proposed strategies are analyzed.
\vspace{-0.3cm}
\subsection{BCSK-d}
In BCSK-d, to ensure the number of absorbed molecules remains at a stable level when a bit 1 is transmitted in the $k$th bit interval, the transmitter adjusts the signal power, namely, the number of molecules, based on the estimated distance. The expected number of molecules $N_{rx}$ in the current bit interval $T$ can be expressed as
\begin{equation}
\small
{N_{rx}}\left( k \right) = {N_{tx}}\left( k \right)\mathrm{erfc}\left( {\frac{{{ \hat{d}_k}}}{{\sqrt {4D_{\mathrm{eff}}T} }}} \right)+ {N_n}\left( k \right),
\end{equation}
where ${N_n}\left( k \right)$ is the additive signal dependent noise and follows a Gaussian distribution ${N_n}\left( k \right)\sim{\mathcal{N}}({\mu _n},\sigma _n^2)$ with mean $\mu_n=0$ and variance $\sigma_n^2$, which is dependent on the expected number of molecules received by the receiver.
For distance $d_k$, to ensure a number $N_{rs}$ of molecules absorbed by the receiver within the current bit interval, we can solve (7) to give the required number of transmitted molecules in the $k$th bit interval:
substituting $\mu_n = 0$ for $N_n(k)$, $N_{tx}(k)$ is
\begin{equation}
\small
{N_{tx}}\left( k \right) = {\raise0.7ex\hbox{${ {{N_{rs}}}}$} \!\mathord{\left/
 {\vphantom {{\left( {{N_{rs}} - {N_n}\left( k \right)} \right)} {erfc\left( {\frac{{{{\hat d}_k}}}{{\sqrt {4{D_{eff}}t} }}} \right)}}}\right.\kern-\nulldelimiterspace}
\!\lower0.7ex\hbox{${\mathrm{erfc}\left( {\frac{{{{\hat d}_k}}}{{\sqrt {4{D_{\mathrm{eff}}}T} }}} \right)}$}}.    
\end{equation}%
As can be seen from (8), for a given number of received molecules, the number of transmitted molecules varies with the distance between the transmitter and the receiver.

\subsection{BCSK-d-RM}
In a mobile MC system, residual molecules propagate in the medium and arrive at the receiver in subsequent time slots, causing ISI. Residual molecules can both be beneficial and harmful to the symbol in question. The residual molecules become harmful to the subsequent symbol when the intended symbol is 0, or, they can also be beneficial for consecutive bits 1 as they can enhance the signal power. In the BCSK-d-RM, we jointly consider distance and residual molecules in the channel to adjust the transmitted molecules.

Assuming a bit 1 is transmitted in the $k$th bit interval, including ISI, the number of molecules absorbed by the receiver in the $k$th bit interval ${N_{rx}}\left( k \right)$ is
%
%
\begin{equation}
\small
{N_{rx}}\left( k \right) = N_{rx}^c\left( k \right) + N_{rx}^p\left( k \right)+ {N_n}\left(k \right),    
\end{equation}%
where $N_{rx}^c\left( k \right)$ is the number of molecules from the $k$th bit interval. $N_n$ is the noise described in (7). For a large number of molecules $N_{tx}$ released by the transmitter at the beginning at the time slot $k$, the number of molecules counted by the receiver in the current time slot can be approximated by a normal distribution ${N_{rx}^c\left( k \right)}\sim \mathcal{N}\left( {{N_{tx}\left( k \right)}P_0,{N_{tx}\left( k \right)}P_0\left( {1 - P_0} \right)} \right) $.
The quantity $N_{rx}^p\left( k \right)$ is the number of molecules received from the previous bit interval and represents ISI.
$N_{rx}^p\left( k \right) = \sum\limits_{j = 1}^{k - 1} {\mathbb{N}_{rx}^p\left( j \right)} $ and
$\mathbb{N}_{rx}^p\left( j \right) \sim \mathcal{N}\left( {{N_{tx}}\left( {k - j} \right)b\left( {k - j} \right){P_j},{N_{tx}}\left( {k - j} \right)b\left( {k - j} \right){P_j}\left( {1 - P_j} \right)} \right)$, $j = 1,2,...,k - 1$.
Where $P_j$ denotes the probability that a molecular transmitted in the bit interval $\left[ {jT,\left( {j + 1} \right)T} \right]$. Therefore, to ensure a stable number of molecules $N_{rs}$ received in the current bit interval, the number of transmitted molecules in the $k th$ bit interval $N_{tx(k)}$ can be expressed similarly to (8) as
\begin{equation}
\small
{N_{tx}}\left( k \right) = {\raise0.7ex\hbox{${\left( {{N_{rs}} -  {N_{rx}^p\left( k \right)}} \right)}$} \!\mathord{\left/
 {\vphantom {{\left( {{N_{rs}} - {N_{rx}^p\left( k \right)}  - {N_n}\left( k \right)} \right)} {erfc\left( {\frac{{{{\hat d}_k}}}{{\sqrt {4{D_{eff}}t} }}} \right)}}}\right.\kern-\nulldelimiterspace}
\!\lower0.7ex\hbox{${\mathrm{erfc}\left( {\frac{{{{\hat d}_k}}}{{\sqrt {4{D_\mathrm{eff}}T} }}} \right)}$}}.   
\end{equation}%
As can be seen from $\left( 10 \right)$, the computational complexity for BCSK-d-RM is related to the length of channel memory. The computational complexity is $o(1)$ and $o( k )$ for BCSK-d and BCSK-d-RM, respectively. Therefore, the computational complexity of BCSK-d is lower than BCSK-d-RM.
\subsection{BER analysis}
%
%
In mobile MC, an error occurs if ${ \hat {b}_k} \ne {b_k}$, where $\hat {b}_k$ denotes the bit received in the $k$th time slot and ${b_k}$ denotes the bit transmitted at the beginning of the time slot. The bit detection at the receiver can be formulated as a binary hypothesis testing problem, where $H_0$ and $H_1$ indicate the transmitter sending bit 0 and bit 1, respectively. The number of molecules received in the $k$th bit interval have distributions, under
  ${H_0}:{N_{rx}}\left( k \right) \sim N\left( {{\mu _0}\left( k \right),\sigma _0^2\left( k \right)} \right)$; and, under $
{H_1}:{N_{rx}}\left( k \right) \sim N\left( {{\mu _1}\left( k \right),\sigma _1^2\left( k \right)} \right)$. 
Here, $\mu_0$ and $\mu_1$ are the mean, and $\sigma_0^2$ and $\sigma_1^2$ are the variance of the received molecules under hypotheses $H_0$ and $H_1$, respectively. These quantities are given by:
\begin{align}
{\mu _0}\left( k \right) &= {\mu _p}\left( k \right) + {\mu _n}\left( k \right) = \frac{1}{2}\sum\limits_{j = 1}^{k - 1} {{N_{tx}}\left( {k - j} \right){P_j}}, \\
\sigma _0^2\left( k \right) &= \sum\limits_{j = 1}^{k - 1} {\sigma _p^2\left( j \right)}  + \sigma _n^2\left( k \right)\\\nonumber
&\hspace{-10mm}= \sum\limits_{j = 1}^{k - 1} {\left[ {\frac{1}{2} {{N_{tx}}\left( {k - j} \right){P_j}\left( {1 - {P_j}} \right)} + \frac{1}{4}{{\left( {{N_{tx}}\left( {k - j} \right){P_j}} \right)}^2}} \right]}  + {\mu _0}\left( k \right)
\end{align}
\begin{align}
{\mu _1}\left( k \right) &= {\mu _c}\left( k \right) + {\mu _p}\left( k \right) + {\mu _n}\left( k \right) \\\nonumber
& = {N_{tx}}\left( k \right){P_0} + \frac{1}{2}\sum\limits_{j = 1}^{k - 1} {{N_{tx}}\left( {k - j} \right){P_j}},\\
 \sigma _1^2\left( k \right) &= {\sigma_c ^2}\left( k \right) + \sum\limits_{j = 1}^{k - 1} {\sigma _p^2\left( j \right)}  + \sigma _n^2\left( k \right)\\\nonumber
 &\hspace{-10mm}= {N_{tx}}\left( k \right){P_0}\left( {1 - {P_0}} \right) + \sum\limits_{j = 1}^{k - 1} {\left[ {\frac{1}{2} {{N_{tx}}\left( {k - j} \right){P_j}\left( {1 - {P_j}} \right)}} \right]}\\\nonumber
 &\hspace{-10mm}+ \sum\limits_{j = 1}^{k - 1} {\left[ {\frac{1}{4}{{\left( {{N_{tx}}\left( {k - j} \right){P_j}} \right)}^2}} \right]}  + {\mu _1}\left( k \right).   
\end{align}

Given ${b_1^{k-1}}$, which means the bit transmitted from $1$ to $k-1$ bit interval, the probability of error $P_e(k)$ for the same probability to transmit bit 0 and bit 1 in the $k$th time slot corresponding to the number of transmitted molecules and the detection threshold, can be written as 
%
\begin{align}
\small
\nonumber
{P_e}\left( {k|{b_1^{k-1}}} \right) &= \frac{1}{2}\Pr \left( {{N_{rx}}(k) < {N_{th}}|{b_k} = 1,{b_1^{k-1}}} \right)\\
\nonumber
&\:\:\:\:+ \frac{1}{2}\Pr \left( {{N_{rx}}(k) > {N_{th}}|{b_k} = 0,{b_1^{k-1}}} \right)\\
&= \frac{1}{2}\left( {1 - {P_D}\left( {k|{b_1^{k-1}}} \right)} \right) + \frac{1}{2}{P_F}\left( {k|{b_1^{k-1}}} \right)
\end{align}
%
where
${P_D}\left( {k|{b_{k-1}}} \right)$ and
${P_F}\left( {k|{b_{k-1}}} \right)$ denote the probability of detection and false alarm at the receiver in the $k$th time slot, respectively. These can be written as 
%
\begin{align}
  {P_D}\left( {k|{b_1^{k-1}}} \right) &= \Pr \left( {{ \hat {b}_k} = 1|{b_k} = 1,{b_1^{k-1}}} \right)\\ \nonumber
  &=  {Q\left( {\frac{{N_{th} - {\mu _1}\left( k \right)}}{{{\sigma _1}\left( k \right)}}} \right)}, \\
  {P_F}\left( {k|{b_1^{k-1}}} \right) &= \Pr \left( {{ \hat {b}_k} = 1|{b_k} = 0,{b_1^{k-1}}} \right)\\ \nonumber
  &=  {Q\left( {\frac{{N_{th} - {\mu _0}\left( k \right)}}{{{\sigma _0}\left( k \right)}}} \right)} 
\end{align}
%
where $N_{th}$ is the detection threshold, and $Q\left(  \cdot  \right)$ is the tail probability of the normal distribution and can be expressed as $Q\left( x \right) = \frac{1}{{\sqrt {2\pi } }}\int_x^\infty  {{e^{ - \frac{{{u^2}}}{2}}}} du$.

%
%
\section{performance evaluation}
In this section, performance evaluations are conducted by Monte Carlo simulations to verify the feasibility of ISI mitigation with the proposed power control schemes. 
In our simulations, $d_0 = 5$mm, $T = 1$ms,  
velocity of the medium is $1$ mm/s, diffusion coefficient of the molecules and the receiver are  $60$ $\mu$m$^2$/s and $10$ $\mu$m$^2$/s, respectively.


In Fig. 2, we analyze the channel impulse response (CIR) that varies with distance $d$. The peak of CIR decreases with distance while the time corresponding to the peak CIR increasing. The decreasing rate is getting smaller and the increasing rate gets larger with the increase of distance.
\begin{figure}[!t]
  \centering
  \includegraphics[width=0.35\textwidth]{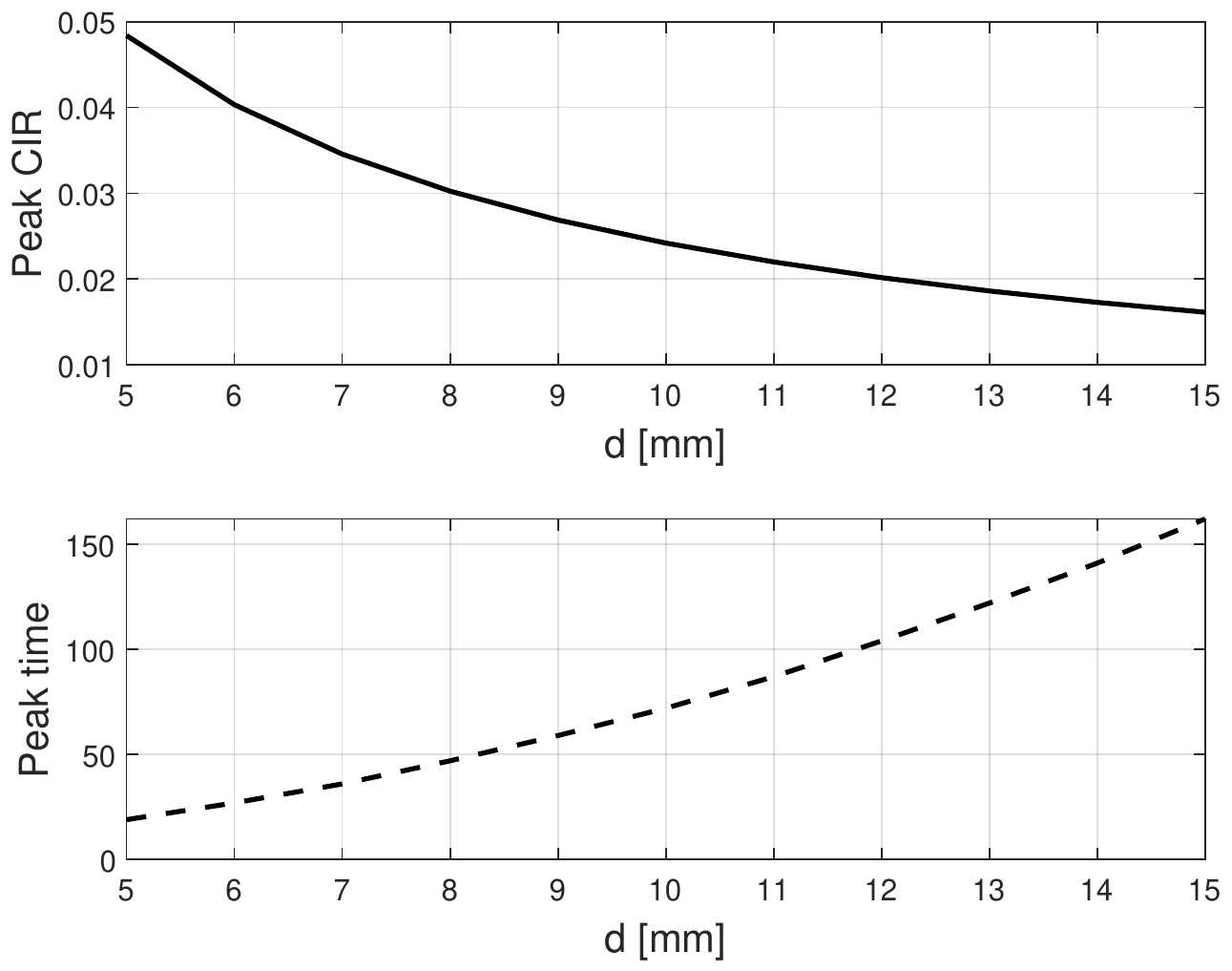}\\
  \caption{The peak CIR and the corresponding time varies with distance $d$.}
\end{figure}

%
%
In Fig. 3, we compare the actual distance (indicated as Ad in the figure), estimation distance (Ed in the figure) and estimation distance-dc (Ed-dc in the figure), for different drift velocities and diffusion coefficients. Estimation distance is obtained from (5) while the estimation distance-dc applies the distance compensation term shown in (6). Without compensation, at larger velocities, the performance of estimation distance is poor because the receiver moves a significant distance during the estimation process; however, compensation restores near-optimal performance. In general, the performance of estimation is degraded as the diffusion coefficient goes up, though the performance of estimation distance-dc is still good. Throughout the remainder of the results, we use estimated distance from (6).

\begin{figure}[!t]
  \centering
  \includegraphics[width=0.35\textwidth]{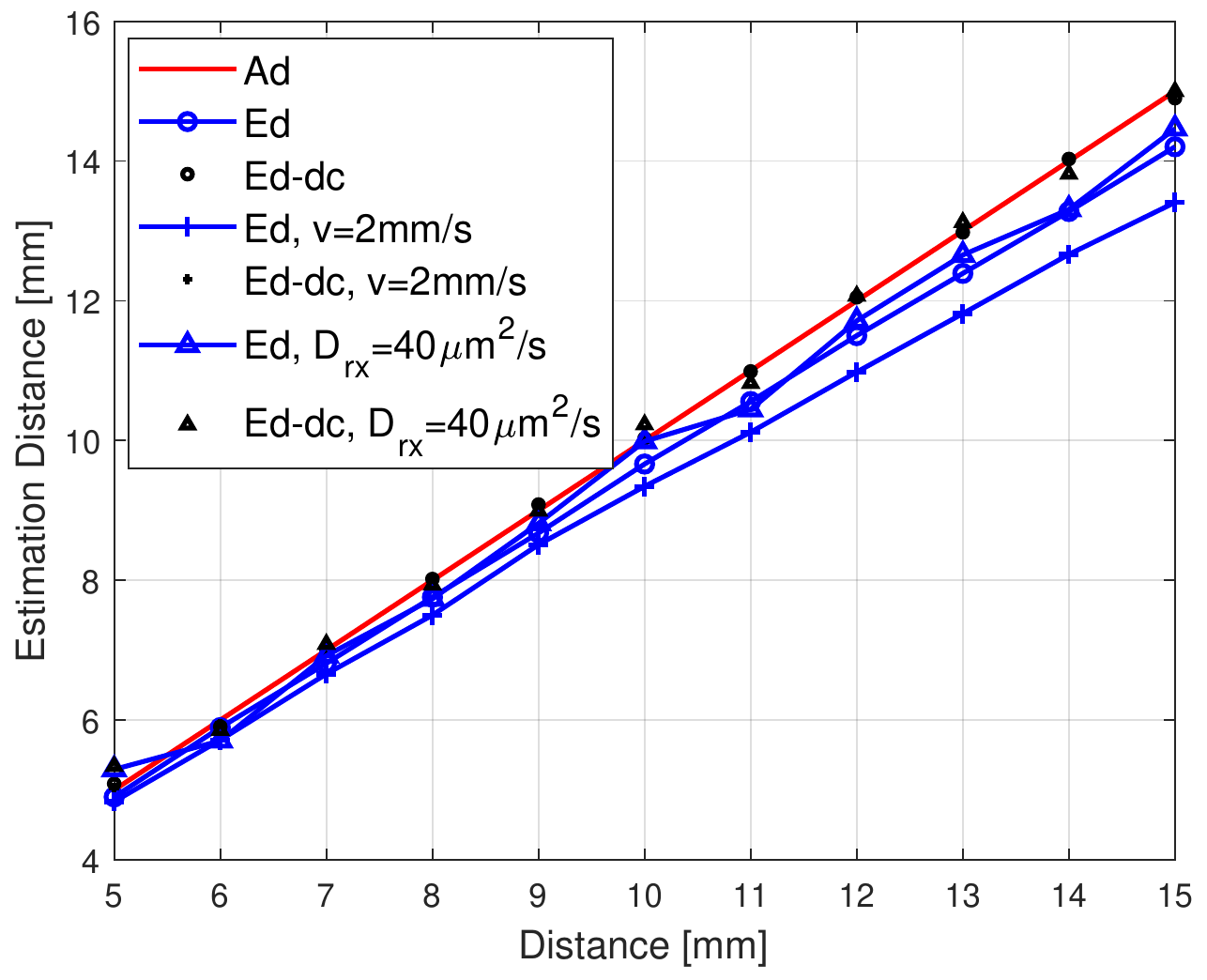}\\
  \caption{Distance estimation performance for different $v$ and $D_{rx}$: Ad represents actual distance, Ed represents estimation distance (from (5)), and Ed-dc represents estimation distance with compensation (from (6)).}
\end{figure}
In Fig. 4, we compare BER of the BCSK, BCSK-d and BCSK-d-RM schemes, and we see that the latter two schemes achieve outstanding performance. Comparing BCSK and BCSK-d-RM, the best performance for BCSK is achieved at a particular distance: this is because, for a given transmission power and detection threshold, the optimal distance is unique. However, as there are residual molecules in the channel affecting the following transmission, the BER performance cannot better than BCSK-d-RM. At a point where the BER performance of BCSK is nearly to the BCSK-d-RM, BCSK uses more molecules. 
Meanwhile, comparing BCSK-d and BCSK-d-RM, the transmitter adjusts the number of molecules according to the measured distance in BCSK-d. 
Thus, BCSK-d-RM outperforms BCSK-d, although the latter has similar performance for lower computational complexity. 
With the increase of distance, the gap between the BER at the actual distance and the estimated distance is decreased, as the error in distance estimation is mostly due to Brownian motion. As distance increases, the estimated distance is a better estimate of the actual distance.
\begin{figure}[!t]
  \centering
  \includegraphics[width=0.35\textwidth]{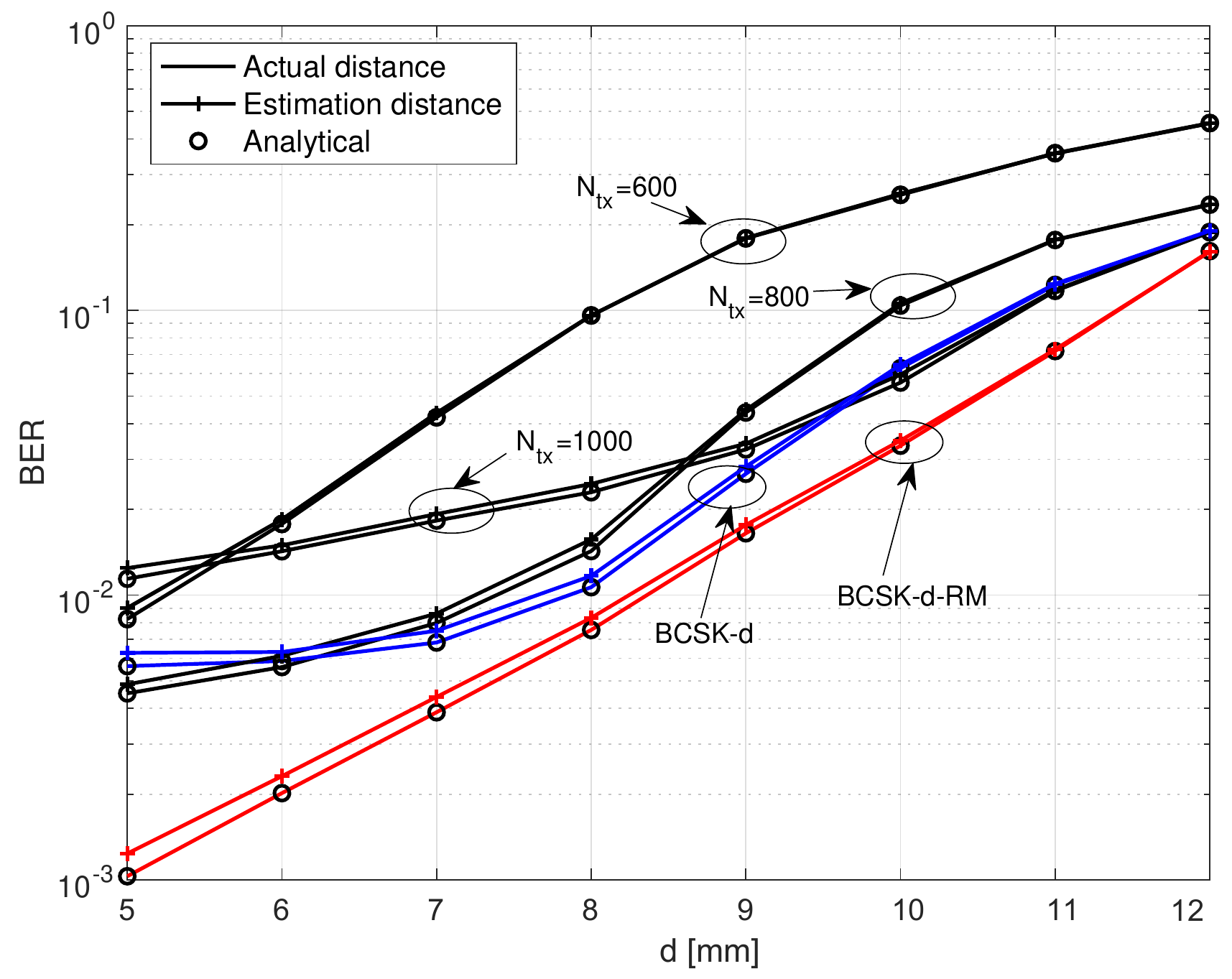}\\
  \caption{The BER varies with distance.}
\end{figure}

In Fig. 5, we compare the number of transmitted molecules with the distance. Using BCSK without power control, the number of transmitted molecules per bit is not varied with the distance. However, with the increase of distance, the number of transmitted molecules per bit increases for BCSK-d and BCSK-d-RM due to power control, while the gap between the BCSK-d and BCSK-d-RM also increases. 
%
%

\begin{figure}[!t]
  \centering
  \includegraphics[width=0.35\textwidth]{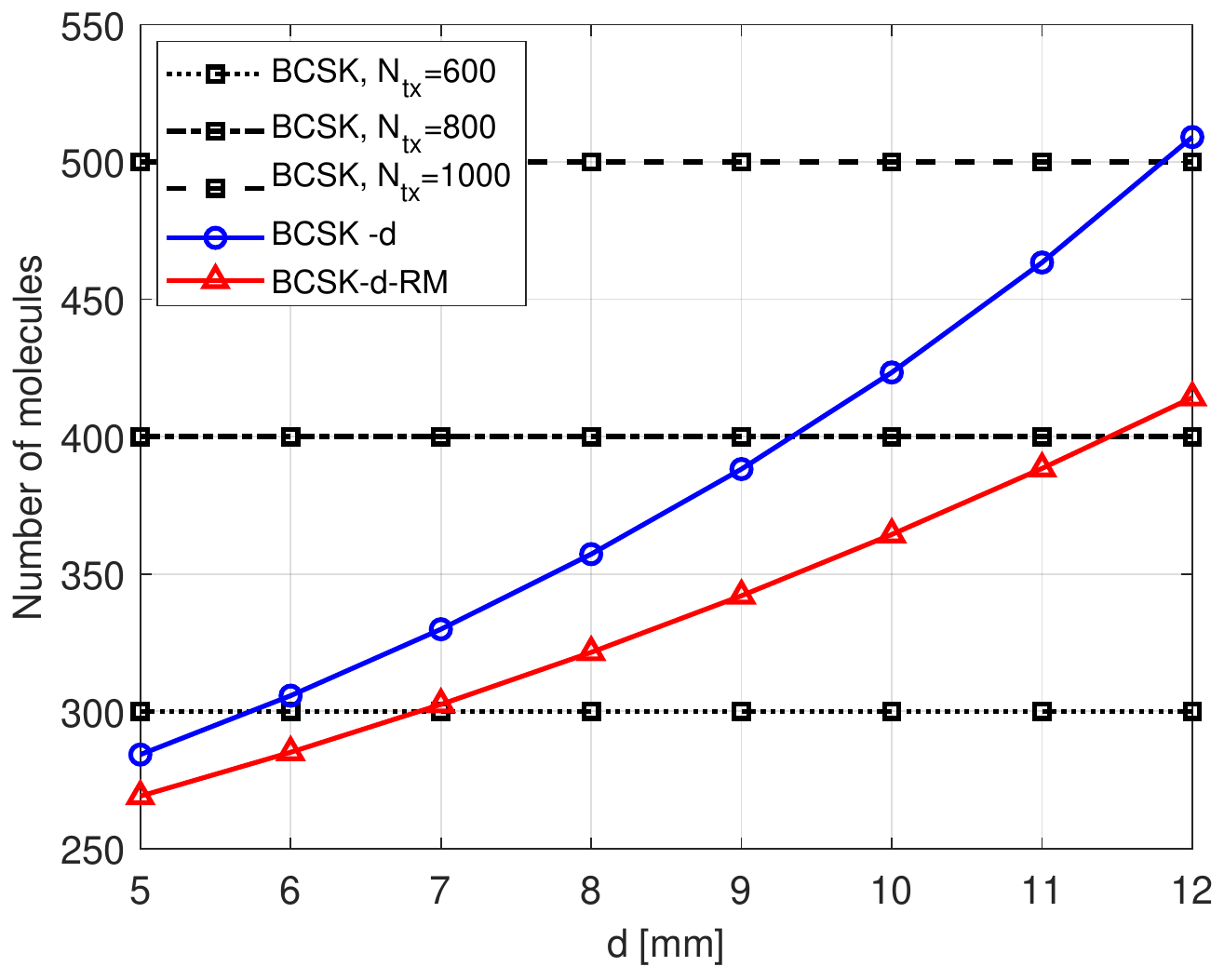}\\
  \caption{The number of transmitted molecules per bit varies with distance.}
\end{figure}

In Fig. 6, we consider the effect of detection threshold on bit error rate as a function of distance.
With the increase of distance, the optimal detection threshold decreases when employing BCSK. This is because the absorbed molecules decrease with the increase of distance. The optimal detection threshold for BCSK-d-RM remains nearly unchanged, but the optimal detection threshold for BCSK-d-RM is slightly smaller than the BCSK-d, as the residual molecules accumulate in the channel.
\begin{figure}[!t]
  \centering
  \includegraphics[width=0.35\textwidth]{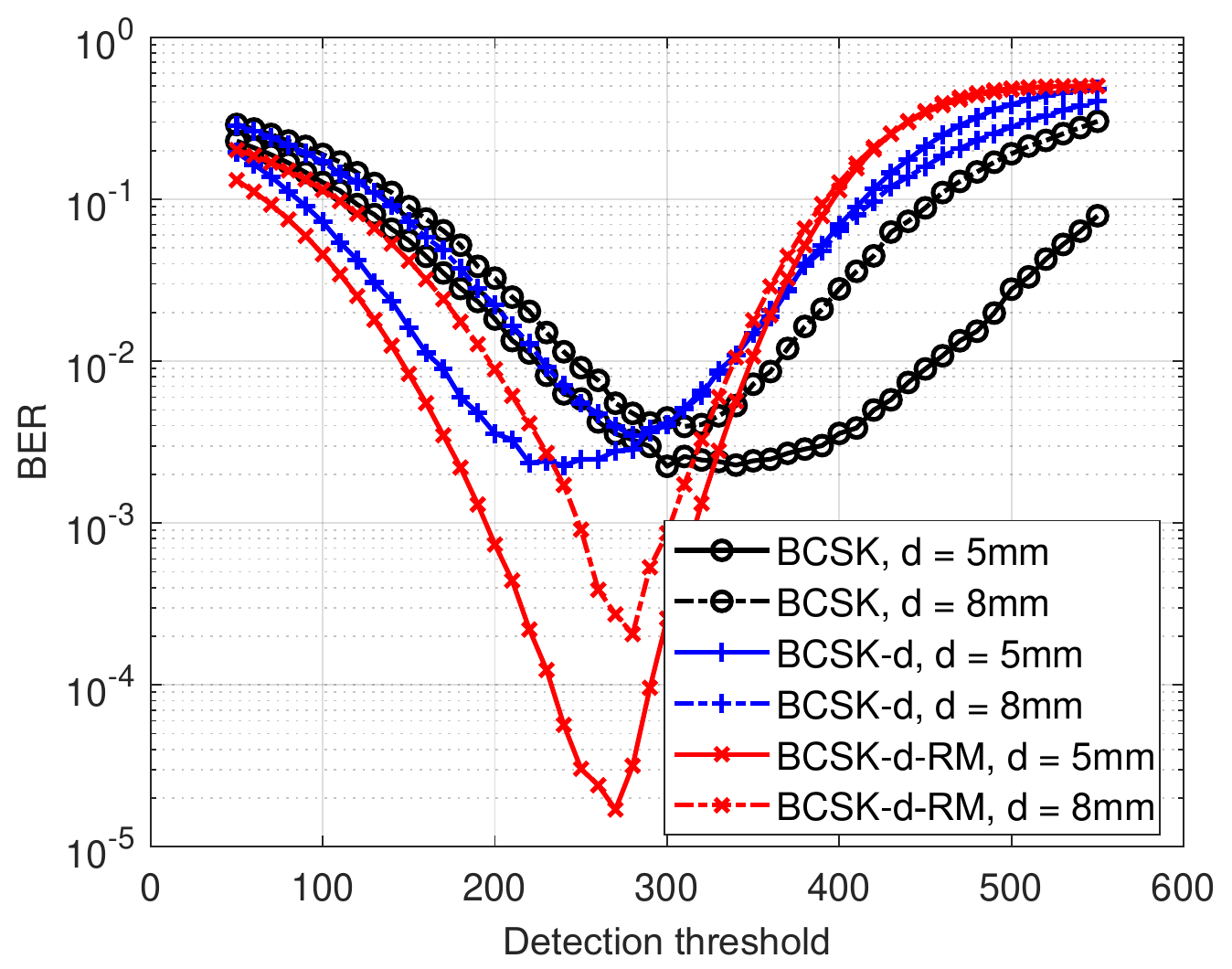}\\
  \caption{The BER varies with the detection threshold.}
\end{figure}

\section{conclusion}
In this letter, we proposed two power control strategies for the mobile MC to improve the communication reliability. 
In BCSK-d, the transmitter adjusts the emitted molecules based on the distance between the transmitter and the receiver, while in BCSK-d-RM, the transmitter jointly considers the distance and residual molecules in the channel. 
The choice between BCSK-d (with lower computational complexity) and BCSK-d-RM (with better performance) gives the option to trade off complexity and performance in a mobile MC system.

\bibliographystyle{IEEEtran}
\bibliography{references}

\begin{thebibliography}{10}
\providecommand{\url}[1]{#1}
\csname url@samestyle\endcsname
\providecommand{\newblock}{\relax}
\providecommand{\bibinfo}[2]{#2}
\providecommand{\BIBentrySTDinterwordspacing}{\spaceskip=0pt\relax}
\providecommand{\BIBentryALTinterwordstretchfactor}{4}
\providecommand{\BIBentryALTinterwordspacing}{\spaceskip=\fontdimen2\font plus
\BIBentryALTinterwordstretchfactor\fontdimen3\font minus
  \fontdimen4\font\relax}
\providecommand{\BIBforeignlanguage}[2]{{%
\expandafter\ifx\csname l@#1\endcsname\relax
\typeout{** WARNING: IEEEtran.bst: No hyphenation pattern has been}%
\typeout{** loaded for the language `#1'. Using the pattern for}%
\typeout{** the default language instead.}%
\else
\language=\csname l@#1\endcsname
\fi
#2}}
\providecommand{\BIBdecl}{\relax}
\BIBdecl

\bibitem{farsad2016comprehensive}
N.~Farsad, H.~B. Yilmaz, A.~Eckford, C.-B. Chae, and W.~Guo, ``A comprehensive
  survey of recent advancements in molecular communication,'' \emph{IEEE
  Communications Surveys \& Tutorials}, vol.~18, no.~3, pp. 1887--1919, 2016.

\bibitem{akyildiz2019moving}
I.~F. Akyildiz, M.~Pierobon, and S.~Balasubramaniam, ``Moving forward with
  molecular communication: from theory to human health applications,''
  \emph{Proceedings of the IEEE}, vol. 107, no.~5, pp. 858--865, 2019.

\bibitem{haselmayr2017transposition}
W.~Haselmayr, S.~M.~H. Aejaz, A.~T. Asyhari, A.~Springer, and W.~Guo,
  ``Transposition errors in diffusion-based mobile molecular communication,''
  \emph{IEEE Communications Letters}, vol.~21, no.~9, pp. 1973--1976, 2017.

\bibitem{jamali2019channel}
V.~Jamali, A.~Ahmadzadeh, W.~Wicke, A.~Noel, and R.~Schober, ``Channel modeling
  for diffusive molecular communication—a tutorial review,''
  \emph{Proceedings of the IEEE}, vol. 107, no.~7, pp. 1256--1301, 2019.

\bibitem{iwasaki2017graph}
S.~Iwasaki and T.~Nakano, ``Graph-based modeling of mobile molecular
  communication systems,'' \emph{IEEE Communications Letters}, vol.~22, no.~2,
  pp. 376--379, 2017.

\bibitem{huang2018mean}
S.~Huang, L.~Lin, H.~Yan, J.~Xu, and F.~Liu, ``Mean and variance of received
  signal in diffusion-based mobile molecular communication,'' in \emph{2018
  IEEE Global Communications Conference (GLOBECOM)}.\hskip 1em plus 0.5em minus
  0.4em\relax IEEE, 2018, pp. 1--6.

\bibitem{varshney2018diffusive}
N.~Varshney, A.~K. Jagannatham, and P.~K. Varshney, ``On diffusive molecular
  communication with mobile nanomachines,'' in \emph{2018 52nd Annual
  Conference on Information Sciences and Systems (CISS)}.\hskip 1em plus 0.5em
  minus 0.4em\relax IEEE, 2018, pp. 1--6.

\bibitem{sun2016adaptive}
Y.~Sun, M.~Ito, and K.~Sezaki, ``Adaptive code width protocol for mitigating
  intersymbol interference in diffusion-based molecular communication with
  mobile nodes,'' in \emph{2016 IEEE 18th International Conference on e-Health
  Networking, Applications and Services (Healthcom)}.\hskip 1em plus 0.5em
  minus 0.4em\relax IEEE, 2016, pp. 1--6.

\bibitem{chang2017adaptive}
G.~Chang, L.~Lin, and H.~Yan, ``Adaptive detection and isi mitigation for
  mobile molecular communication,'' \emph{IEEE Transactions on nanobioscience},
  vol.~17, no.~1, pp. 21--35, 2017.

\bibitem{tepekule2015isi}
B.~Tepekule, A.~E. Pusane, H.~B. Yilmaz, C.-B. Chae, and T.~Tugcu, ``Isi
  mitigation techniques in molecular communication,'' \emph{IEEE Transactions
  on Molecular, Biological and Multi-Scale Communications}, vol.~1, no.~2, pp.
  202--216, 2015.

\bibitem{varshney2018flow}
N.~Varshney, W.~Haselmayr, and W.~Guo, ``On flow-induced diffusive mobile
  molecular communication: First hitting time and performance analysis,''
  \emph{IEEE Transactions on Molecular, Biological and Multi-Scale
  Communications}, vol.~4, no.~4, pp. 195--207, 2018.

\bibitem{moore2012measuring}
M.~J. Moore, T.~Nakano, A.~Enomoto, and T.~Suda, ``Measuring distance from
  single spike feedback signals in molecular communication,'' \emph{IEEE
  Transactions on Signal Processing}, vol.~60, no.~7, pp. 3576--3587, 2012.

\bibitem{kim1999exact}
H.~Kim and K.~J. Shin, ``Exact solution of the reversible diffusion-influenced
  reaction for an isolated pair in three dimensions,'' \emph{Physical review
  letters}, vol.~82, no.~7, p. 1578, 1999.

\end{thebibliography}
\end{document}